\documentclass[a4paper,11pt,onecolumn]{article}
\usepackage{amsmath}
\usepackage{amsfonts}
\usepackage{amssymb}
\usepackage{graphicx}
\usepackage{bm}
\usepackage{geometry}
\usepackage{cite}
\usepackage[makeroom]{cancel}
\usepackage[usenames,dvipsnames]{color}
\usepackage[plainpages=false,hypertexnames=true,hyperindex=true,pagebackref=false,raiselinks=true]{hyperref}

\renewcommand{\vec}{\bm}

\begin{document}

\title{Contextuality of the probability current in quantum mechanics}
\author{F. Lalo\"{e}\thanks{%
laloe@lkb.ens.fr} \\
Laboratoire Kastler Brossel, ENS-Universit\'e PSL,\\
CNRS, Sorbonne Universit\'e, Coll\`ege de France,\\
24 rue Lhomond 75005\ Paris, France}
\date{\today }
\maketitle

\begin{abstract}
We revisit an argument proposed by Hardy
\cite{Hardy-1} concerning local realistic theories,
but  in terms of the motion of the probability fluid
 and its
current within
standard quantum mechanics.
For a two-particle system, we discuss surprising properties  of the flux lines
of this current in configuration space, in particular
(quasi)
discontinuous variations  when the context (the
experimental setup) is changed. This occurs in both
Galilean and Einsteinian relativity. In the latter
case, discontinuous variations also appear without
changing the setup, when the Lorentzian reference frame
is changed. A relativistically consistent definition of the motion
of the fluid of quantum probability in configuration space is therefore impossible in this case.
\end{abstract}


Most textbooks in
quantum mechanics introduce, in one of their early chapters, a probability current
 $\vec J (\vec r )$ when discussing
the motion of a single particle in 3-dimensional space. This current
 defines the motion of a probability fluid,
 which provides a useful picture of the quantum evolution.
For a system of $N$ particles, a straightforward
generalization provides a probability
current $\vec J (\vec r_1 , \vec r_2 , \ldots , \vec r_N )$
defined in configuration space (the space associated with
the $N$ spatial coordinates
$\vec r_1 , \vec r_2 , \ldots , \vec r_N$).
The purpose of this article is to emphasize some
surprising properties
of this probability current in relativistic quantum
mechanics.

Within Galilean relativity, the
Schr\"{o}dinger equation directly implies the existence of a
local conservation law of the probability in terms of
a current
$\vec J (\vec r _1, \vec r _2,\ldots , \vec r _N, )$.
One can then build lines of flux,
or trajectories in configuration space, which
are at each point tangent to $\vec J$.
When a different Galilean reference frame is used,
the components of $\vec J$ are transformed as those of
the current of a classical fluid moving in configuration space. It is then possible
to develop pictorial views of the motion of the
probability fluid and, if one wishes to introduce
additional variables in quantum mechanics, to build
the dBB (de Broglie-Bohm) theory
\cite{de-Broglie, Bohm, Holland}
by adding positions variables. Nevertheless, in this article, we remain
within standard quantum mechanics and limit ourselves to
the properties of
its probability current.

Within Einsteinian relativity, we shall see that
the situation is more complex since, for a system of several particles, it is not possible
to define a probability current that takes consistent
values in different Lorentz frames. The current therefore depends on the reference
frame used by the observer and may be called contextual, by a slight
generalization of the usual meaning of the word
\textquotedblleft contextual\textquotedblright\
in quantum mechanics.

The essence of our discussion is very similar to that
already given by  Hardy
\cite{Hardy-1}, but here we  emphasize the motion of the
probability fluid and the role of the probability current,
instead of the EPR \textquotedblleft elements of
reality\textquotedblright\ \cite{EPR}. We feel that this
pint of view  may be useful, since most physicists are more
familiar with the notion of probability current (often used in collision theory
for instance) than with discussions on the foundations
of physics. In \S~\ref{galilean},
we first analyze the experiment within Galilean relativity. We show that, even in this simple case,
the probability current in configuration space  exhibits peculiar contextual
properties (dependence on the whole experimental setup); in particular, it can change abruptly
when the experimental conditions (the context)  are
slightly modified. In \S~\ref{einsteinian} we
discuss the same experiment within Einsteinian relativity,
and show that the properties of the probability current
are even more surprising: when the same
experiment is described in a different Lorentz reference
frame, the probability current may look very
different; it may even exhibit (almost) discontinuous
variations that have very little to do with those  of the
current of a classical fluid. This puts severe
constraints on the construction of theories with
additional variables, as emphasized in \cite{DGNSZ}.

Somewhat related considerations have been developed
by Gisin
\cite{Gisin}, who emphasizes that even nonlocal
deterministic theories cannot reproduce the results
of quantum mechanics while remaining covariant. Berndl et al.
\cite{Berndl-et-al} propose a discussion that is also
similar to ours; they put the emphasis on the absence of
relativistic quantum equilibrium and on the possibilities
offered by a multitime dBB formalism, leading to the introduction
of synchronized $N$-path for $N$ noninteracting
particles. Here we remain
within the scope of standard quantum mechanics to discuss,
for spinless particles, peculiar properties of the
probability current.

\section{Galilean relativity}\label{galilean}

\begin{figure}[h]
\centering
\includegraphics[trim = 60mm 55mm 65mm 40mm, clip,
width=9cm]{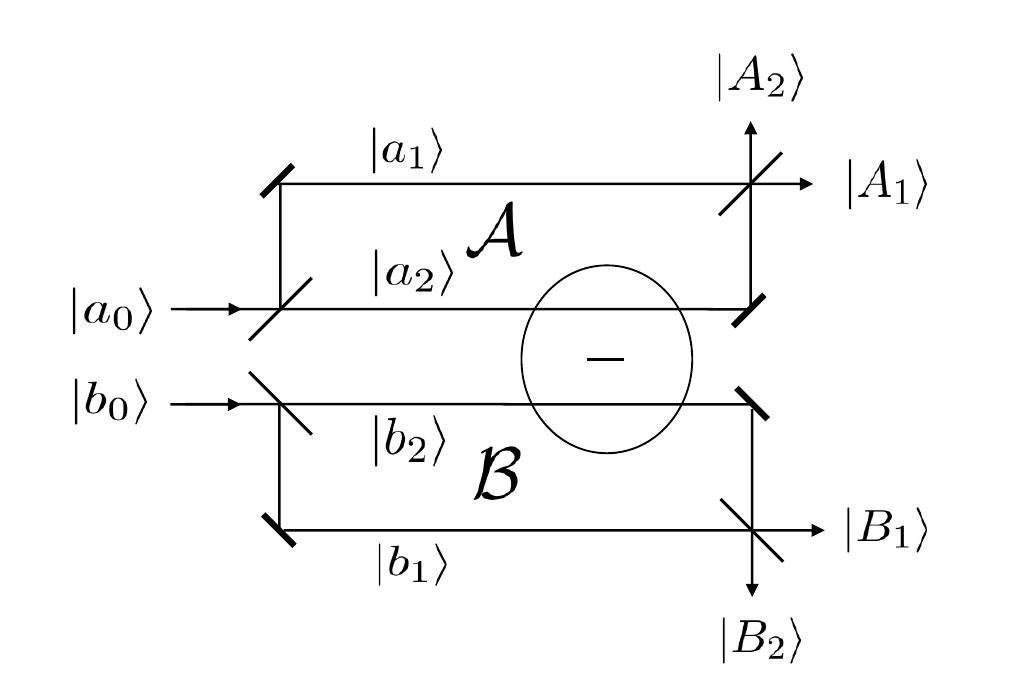}
\caption{Each of the two interferometers $\mathcal A$
and $\mathcal B$
receives a particle. The oval symbolizes an interaction
process that annihilates the two particles if they both
take
the middle paths. To survive, the system of two particles
has to propagate in the other combinations of paths, which results in
quantum entanglement.}
\label{f1}
\end{figure}

Following Hardy \cite{Hardy-1}, we consider two
interferometers $\mathcal A $ and $\mathcal B$
(figure~\ref{f1}). If the particles fly across
the arms $a_2$ and $b_2$ in the middle of the figure,
they interact and annihilate each other by some process.
The two-particle system, if it survives,
can only propagate in the three other paths
$a_1 b_1$, $a_2 b_1$ , or $a_1 b_2$, and  becomes entangled.

We denote by $\vert a_0 \rangle$ and $\vert b_0
\rangle$ the quantum states of the particles entering the
interferometers. These states can be seen,
either as plane waves
of well defined energy-momentum, or, in a second step,
as wave packets propagating in time and obtained by
linearly
combining these plane waves (writing wave packets
explicitly would make the equation more complicated without
changing anything in the reasoning). Similarly, we denote by
$\vert a_1 \rangle$ and $\vert a_2
\rangle$ (first particle), $\vert b_1 \rangle$ and
$\vert b_2 \rangle$ (second particle), their quantum
states when they propagate inside the interferometers,
as indicated in  figure~\ref{f1}. Finally $\vert A_1
\rangle$ and $\vert A_2 \rangle$, $\vert B_1 \rangle$
and $\vert B_2 \rangle$ are the quantum states of each
particle after it crosses the output beam splitter.

The effect of these output beam splitters is described
by the substitutions
\cite{Garrison-Chiao}:
\begin{equation}\label{changement-kets}
\vert a_1 \rangle \Rightarrow \frac{1}{\sqrt 2} (\vert A_1 \rangle +i \vert A_2 \rangle )
\hspace{2cm}
\vert a_2 \rangle \Rightarrow \frac{1}{\sqrt 2} (i\vert A_1 \rangle + \vert A_2 \rangle )
\end{equation}
and similar substitutions for the interferometer
$\mathcal B$, as well as for the input beam splitters.

\subsection{State vectors}\label{par-1}

Just after crossing the two first beam splitters,
the system is in state:
\begin{equation}\label{1}
 \vert a_0 \rangle \vert b_0 \rangle
\Rightarrow
\frac{1}{2} \left[ i \vert a_1 \rangle + \vert a_2 \rangle  \right]
\left[ i \vert b_1 \rangle + \vert b_2 \rangle  \right]
\end{equation}
Suppressing the $\vert a_2 \rangle \vert b_2 \rangle$ component leads to the entangled state of the two particles:
\begin{equation}\label{2}
\vert \psi \rangle =
\frac{1}{\sqrt 3} \left[ - \vert a_1 \rangle \vert b_1 \rangle +i \vert a_1 \rangle \vert b_2 \rangle
+ i \vert a_2 \rangle \vert b_1 \rangle   \right]
\end{equation}
(since we are interested only in systems that have survived the
annihilation process, we have normalized this
new state to unity).

If one extends the arms of interferometer $\mathcal B$ a shown in figure~\ref{f2}, and if one is interested in the state of the system before the second particle reaches its output beam splitter (or if this beam splitter is removed), one can use relation (\ref{changement-kets}) to obtain the state :
\begin{subequations}\label{eqns-4}
\begin{equation}\label{4-a}
 \vert \psi_A ' \rangle
=
\frac{1}{\sqrt 6} \left[-  2 \, \vert A_1 \rangle \vert b_1 \rangle + i \, \vert A_1 \rangle \vert b_2 \rangle -   \, \vert A_2 \rangle \vert b_2 \rangle   \right]
\end{equation}
We note the absence of a component in
$\vert A_2 \rangle \vert b_1 \rangle$. If, instead,
we consider the situation of  figure~\ref{f3} and take
into account only the output beam splitter of
interferometer $\mathcal A$, we obtain the state:
\begin{equation}\label{4-b}
 \vert \psi_B ' \rangle
=
\frac{1}{\sqrt 6} \left[- 2 \, \vert a_1 \rangle \vert B_1 \rangle + i \, \vert a_2 \rangle \vert B_1 \rangle -   \vert a_2 \rangle \vert B_2 \rangle   \right]
\end{equation}
\end{subequations}
from which the component in $\vert a_1 \rangle \vert B_2 \rangle$ has now disappeared.

Finally, when both output beam splitters are     taken into account, the system is described by the ket:
\begin{equation}\label{5}
\vert \psi '' \rangle =
\frac{1}{\sqrt{12}} \Big[- 3 \, \vert A_1 \rangle \vert B_1 \rangle - i \, \vert A_1 \rangle \vert B_2 \rangle - i  \vert A_2 \rangle \vert B_1 \rangle
- \vert A_2 \rangle \vert B_2 \rangle
 \Big]
\end{equation}
which maintains a non-zero component on all possible
output channels. This is a consequence of the suppression of
the component in $\vert a_2 \rangle \vert b_2 \rangle$ in
(\ref{2}):  if it is retained, applying relation (\ref{1}) twice
to this state leads to the simple relation:
\begin{equation}\label{6}
 \vert a_0 \rangle \vert b_0 \rangle
\Rightarrow i \vert A_1 \rangle \vert B_1 \rangle
\end{equation}
Therefore, if the particles do not interact,
they all exit through channel
$\vert A_1 \rangle \vert B_1 \rangle$.

\subsection{Probability currents and bi-trajectories}

We now consider situations where two particles, described
by wave packets, enter simultaneously the two
interferometers $\mathcal A$ and $\mathcal B$.
The probability current in configuration space
$\vec J (\vec r _1 , \vec r_2 )$
associated with a Schr\"{o}dinger
wave function $\psi (\vec r _1 , \vec r_2 )$
is\footnote{We assume here
that the particles have a non-zero mass, as is the case
for atomic interferometry experiments (in this case, beam splitters can
be obtained with non-resonant lasers and interference
gratings). If the particles are polarized photons, a
different expression
of the current should be used.}:
\begin{equation}\label{7}
  \vec J (\vec r _1 , \vec r_2 )
  = \frac{\hbar}{2mi}\left[ \psi^* (\vec r _1 , \vec r_2 )
   \bm\nabla \psi (\vec r _1 , \vec r_2 )
   - \text{c.c.}   \right]
\end{equation}
where c.c. indicates a complex conjugate. The nabla
derivation operator
has 6 components
in configuration space (3 containing the derivatives
of the wave function with respect to the coordinates
of the first particle, 3 those for the other particle),
and provides the direction along which the pair of
the two wave packets moves in this space. One can then
build continuous
current lines (or flux lines)  in configuration space that
 are
tangent to $\vec J$ at each of their points. Since they are nothing but
pairs of lines in usual 3D space, and since in dBB theory each of these
lines correspond to a trajectory, we will call
them \textquotedblleft bi-trajectories\textquotedblright
. Nevertheless our discussion remains at a purely wave
description,
without introducing Bohmian positions of the two particles
(the variables $\vec r _1$ and $\vec r_2$ are just coordinates labelling the
points in configuration space), so that
this is just a convenient convention of language.

In general (if the wave function is a not product of
single particle wave functions), the current is not a product
of functions of
$\vec r _1$ and $\vec r_2$; the 3D current line in
each interferometer depends on the current line in the other.
We note that the properties of bi-trajectories may look
significantly different from those of a 3D current line for a single
particle. While in the latter case, a 3D no-crossing rule
limits the possible patterns \cite{Holland}, for bi-trajectories the
no-crossing rule applies only in the configuration
space, so that crossings in ordinary space are indeed
possible. For instance, several different current lines may
start from the same point in interferometer $\mathcal A$,
provided they are associated with different lines
in the other interferometer. Similarly, the direction of
each 3D current line may be affected by the insertion of a
beam splitter recombining two wave functions in the other
interferometer. Of course this mutual dependence
occurs only in the presence of entanglement, which
introduces
quantum nonlocal effects.

With wave packets, the probabilities of two-particle measurement
results can be obtained as the time integral of the 6D
flux of the probability current through the input surface
of a pair of detectors\footnote{This operation should be
performed only with the values
of the current at a given time; one should not try to use the
values of the unperturbed probability current at different
times to obtain correlation functions for measurement at
different times. This is related to the effect of the first
measurement on the state vector,
which changes the
probability current in both interferometers (even if
the measurement is performed in only one of them). In
Bohmian theory for instance, the calculation of time
correlation functions requires that the perturbation
of the first measurement on the evolution of the Bohmian
positions is included, before the probability of the
second measurement is calculated
\cite{DWRUQM}.}.
Inside the interferometers, the possible bi-trajectories
associated with the setup shown in figure~\ref{f1} are
obtained from the state vector of equation (\ref{2}),
while equation (\ref{5}) determines them
 after the particles have crossed the output
beam splitters. No bi-trajectory follows the path
$a_2 b_2$, since the corresponding component of the state
vector vanishes: if one 3D current line takes the inner path,
the other has to take the outer path. Whatever the
detailed configurations of the bi-trajectories is, relation
(\ref{5}) indicates that they always reach the 4 output channels,
since they indicate the existence of a
 probability current that brings the
probability into these regions.

\subsection{Different configurations}\label{different-configurations}

\begin{figure}[h]
\centering
\includegraphics[trim = 60mm 50mm 30mm 30mm, clip,width=10cm]{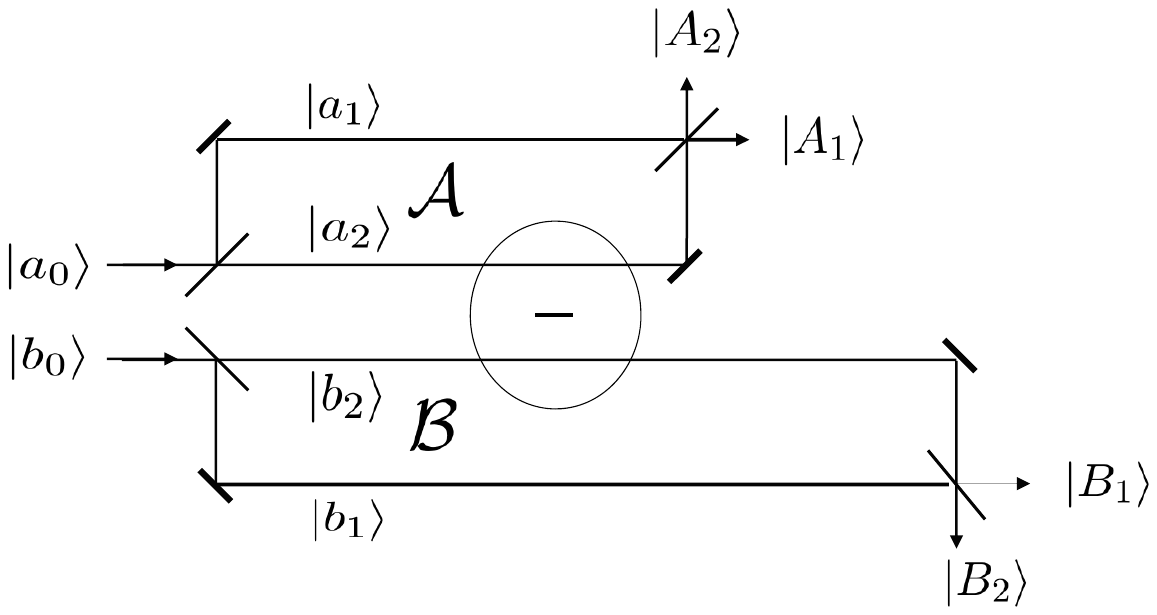}
\caption{A different version of the experiment where the
crossing of the output beam splitter occurs first in
interferometer $\mathcal A$. The lines of the probability
current in configuration space (bi-trajectories) are then
modified in this interferometer. Moreover, if the
bi-trajectory follows $b_1$ in the lower interferometer,
it is certain that it goes to $A_1$ in the upper
interferometer. In other words, the only
bi-trajectories that may exit through $A_2$ must also pass
through $b_2$ .}
\label{f2}
\end{figure}

Now, consider the situation shown in figure~\ref{f2},
where the output of beam splitter of $\mathcal B$  has been
pushed to the right. The particle in interferometer
$\mathcal A$ then crosses its output beam splitter
before the other; in the intermediate time, the state
of the system is given by relation (\ref{4-a}).
This expression  shows that, if the bi-trajectory is such
that the particle in interferometer $\mathcal B$ follows
path $b_1$, the particle in interferometer $\mathcal A$
necessarily exits through $A_1$. The only
bi-trajectories where one particle reaches $A_2$ must have the
other following path $b_2$.


\begin{figure}[h  !]
\centering
\includegraphics[trim = 60mm 50mm 30mm 30mm, clip,width=10cm]{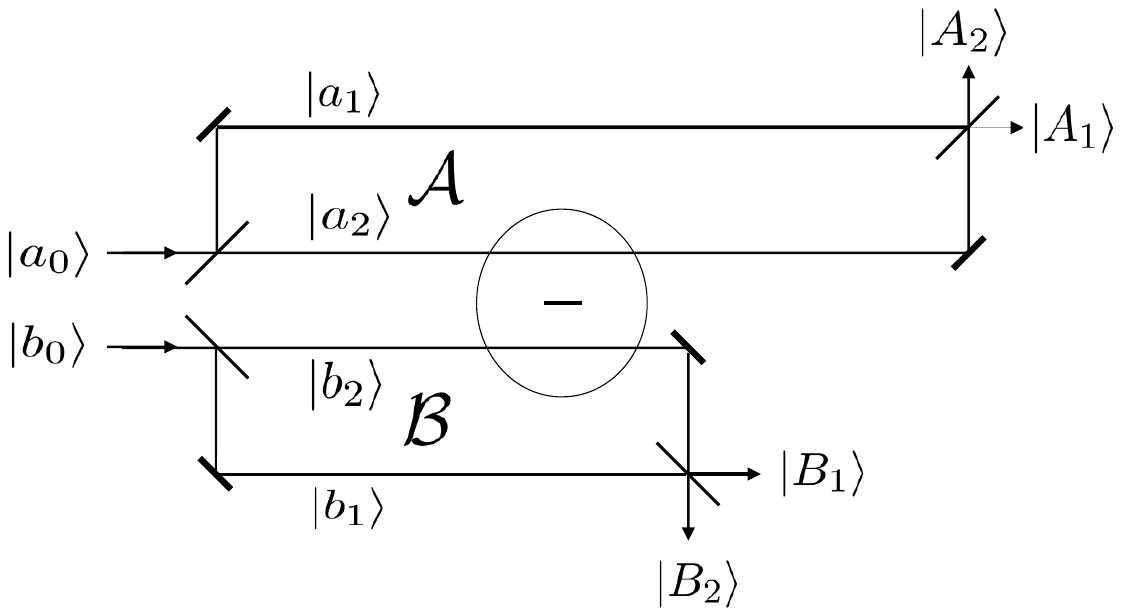}
\caption{A third version of the experiment where the
crossing of the output beam splitter occurs first in
interferometer $\mathcal B$. After this crossing, if
the bi-trajectory follows path $a_1$ in interferometer
$\mathcal A$, it is certain that the other component
reaches $B_1$. The only bi-trajectories that may reach $B_2$
must also pass through $a_2$.}
\label{f3}
\end{figure}

In figure~\ref{f3}, we consider the symmetric situation
where it is the beam splitter of interferometer
$\mathcal A$  that has been pushed to the right.
The state of the system is given by relation (\ref{4-b})
and, in this case, the only bi-trajectories that may reach
exit $B_2$ have to pass through $a_2$.

\subsection{Discussion}

According to relation (\ref{2}) the two particles
cannot be found simultaneously in  $a_2$ and $b_2$,
which shows that no bi-trajectory takes this path.
Relation (\ref{5}) indicates that some bi-trajectories
reach $A_2$ and $B_2$ (more precisely, 1/12 of them). But, in \S~\ref{different-configurations},
we concluded in one case that the only bi-trajectories
reaching $A_2$ have to pass trough $b_2$, and in the
other case that the only bi-trajectories reaching $B_2$
have to pass through $a_2$; combined together, these
two statements seem to indicate that the bi-trajectories
reaching  $A_2$ and $B_2$ have to pass through $a_2$
and $b_2$, which is impossible as we just saw.

This apparent contradiction can be lifted by remarking
that we have combined statements that are valid under
incompatible experimental configurations, those illustrated
in figures~\ref{f2} and \ref{f3}. As a consequence,
there is no reason why the  probability current in configuration space
should
have the same configuration in both cases. It is actually context
dependent, and what matters is the position of beam
splitters and mirrors in the whole apparatus.

The mechanism behind this sensitivity of the probability
current to the whole experimental setup is nonlocal in ordinary
3D space.
When the particle in $\mathcal B$ crosses
its beam splitter, components of the wave function that
were
spatially disjoint (in terms of the positions of that
particle) are made to overlap, which introduces
interference
terms for the wave function guiding the probability current line
in $\mathcal A$.
This nonlocal effect
is directly seen on the expressions (\ref{4-a})
to (\ref{5}) of the quantum state of the physical system.



A striking illustration of this property of the
probability current  in 6D is the appearance of discontinuities
when the experimental situations are modified;
more precisely, very rapid changes of the current lines
occur when one element of the
experimental device is only slightly changed.
When the position of one beam splitter is moved in order
to progressively change the setup from that of
figure~\ref{f2} to that of figure~\ref{f3},  the
probability current lines undergo a sudden variation
when the situation of figure~\ref{f2} is crossed.
Initially, all bi-trajectories reaching $A_2$ and $B_2$
pass through $a_1$ and $b_2$, while at the crossing they
jump discontinuously to $a_2$ and $b_1$ (in dBB theory,
the evolution of the Bohmian position necessarily exhibits
 the same discontinuity).

More precisely, while  wave packets of finite length $\ell$
  are crossing the output beam splitters, the state vector
  may actually be a superposition of states of the form
  (\ref{eqns-4}) and (\ref{5}). The rapid
 change of the pattern of the current lines
 should then take place when the length $L$ of one interferometer
 varies by
 $\ell$;
in fact no real mathematical discontinuity should
then occur, but just
a very rapid change between two regimes of the current
lines for a small variation of $\ell$. At the scale of
$L$, which may be arbitrarily large with respect to $\ell$, the
crossover may appear as a quasi-discontinuity.

Hardy has also discussed this situation in terms
of \textquotedblleft empty waves\textquotedblright\ and their
nonlocal effects \cite{Hardy-2}. In any
theory where localized particles are guided by a wave,
the very possibility of observing pairs of particles
exiting
through outputs $A_2$ and $B_2$ is indeed surprising:
on one hand,
such an event
is possible only if the particles interact by taking the paths
$a_2$ and $b_2$, as seen by comparing relations
 (\ref{5}) and  (\ref{6}); but, on the other hand,
 they are supposed to
disappear if they follow this bi-trajectory !

Ref. \cite{Hardy-3} discusses another version of
the experiment,
where the oval in figure~\ref{f1} symbolizes a dephasing
interaction, instead of the mutual annihilation of the
two particles. This variant of the experiment may be relevant in the context
of the detection of quantum effects induced by
gravity \cite{Bose}.
Simple calculations show that the two versions
of the experiment,
annihilation or dephasing, lead to
very similar conclusions.

\section{Einsteinian relativity}\label{einsteinian}

We now discuss how, in special relativity,  the
sensitivity to the context induces a sensitivity
to the Lorentzian reference frame, preventing a
 relativistically consistent definition of the probability current
in configuration space of the two particle system.
 We first specify
the properties that we assume to be valid within the quantum theory
under discussion.


\subsection{General properties}

Within a relativistic quantum theory, the measurement
results (indications of the pointers of the  apparatuses)
 and their probabilities are relativistic scalars.
 If for
 instance a Stern-Gerlach experiment is observed from
 different Lorentz frames, even if the various observers
 see a different motion of the atoms and of the magnet,
 they all agree that the possible measurement results
 are $\pm \hbar / 2$, as well as on the values of their
 probabilities.  This also applies to correlation
 measurements, provided of course the space-time
 coordinates of the events are
 properly modified.

Within a relativistic quantum theory, we consider a
physical system
made of a fixed number $N$ of massive particles;
we assume that the theory contains
a  local conservation law of the probability,
involving a $N$-particle probability current
defined in configuration space. For example
Refs.
\cite{DGMZ} and \cite{DGNSZ}) propose expressions
of the probability current for relativistic Dirac particles
that may be entangled but do not interact,
which is sufficient for our discussion. Of course, the current takes different
values in different Lorentz frames but, in each frame,
a local law of conservation of the probability can be written
with its own current. This current is necessarily related to the
probabilities of measurement results since, with wave
packets, the flux of this
current (integrated over time) through appropriate
surfaces  (for instance the
input surfaces of detectors) allows one to calculate
probabilities, i.e. relativistic invariants.

For a system of two
particles, the flux has to be calculated at a given time
through a surface in a 6D space, in other words through
a pair or surfaces in ordinary 3D space.
As above, the probability current $\vec J$ allows one to
define
bi-trajectories, which are tangent to $\vec J$ at each
point of configuration space. One can then obtain
the probabilities of measurement results
by making
statistics over the bi-trajectories, which are nothing but statistics over
their starting point (with a weight given by the initial
quantum probability
of presence). This
provides a view of the
theory where the random character does not appear during
measurements, but only in the value of the initial
positions of the bi-trajectories.

The current $\vec J$ can be used in each Lorentzian
frame to obtain the probability of any measurement
performed at a given time in this frame. We will therefore
use it for simultaneous measurements performed with two
detectors sitting at any place in each interferometer (before or after the beam splitters), but not to
derive correlations between results of measurements
performed at different times --  since this would require
to take into account the perturbation of the first
measurement on the current \cite{DWRUQM}.
This necessity can be seen as a consequence of the
von Neumann projection postulate,  or of the entanglement
of the system with the measurement apparatus (for instance,
within dBB theory, the Bohmian positions within this
apparatus
play an important role \cite{Tastevin-Laloe}).

Simultaneous measurements in one Lorentz frame appear
as time shifted measurements in another frame. We will
deduce the
properties of the probability current in each frame
from
probabilities of simultaneous measurements in the same
frame, and then use the fact that probabilities are
relativistic scalar to obtain them
 from calculations in a
different frame.

\subsection{Different views in different Lorentzian frames}

\begin{figure}
\centering
\includegraphics[trim = 60mm 30mm 30mm 20mm, clip,width=8cm]{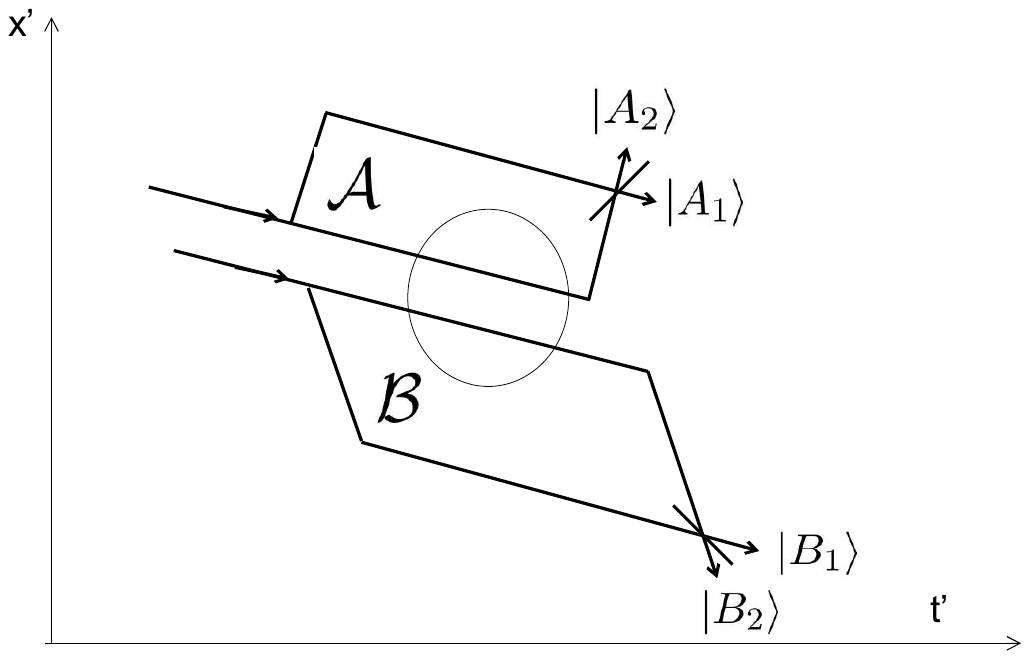}
\caption{This figure shows the experiment of
figure~\ref{f1} in another
 Lorentz frame, moving with a positive velocity $v$ in
 the direction
of the vertical $Ox$ axis in the figure. The particles are
assumed to have a non-zero mass. In interferometer
$\mathcal A$, in some sections the propagation is parallel
to the $Ox$ axis, which increases the velocity; in
interferometer $\mathcal B$ it is
antiparallel, which decreases the velocity.
As a consequence,  the crossing of the
output beam splitter occurs first in interferometer $\mathcal A$
as in figure \ref{f2}. }
\label{f4}
\end{figure}

We now consider a double interference experiment which,
in the Lorentz frame $\mathcal L _0$ of the laboratory,
corresponds to
the scheme shown in figure~\ref{f1}. In another Lorentz
frame,
the same experiment is seen
differently. For instance, figure~\ref{f4} shows a
time-space diagram
describing the experiment in a frame  $\mathcal L _1$
moving
with a positive velocity $v$  along axis $Ox$. The output
beam
splitter is then crossed in interferometer $\mathcal A$
 before it is in interferometer $\mathcal B$. Thus,
 in this Lorentz
frame, it makes sense to consider a simultaneous
measurement of the output
channel in interferometer $\mathcal A$ while the other
particle is still in  interferometer $\mathcal B$; the
probability
of any result can then be obtained by time integration of
the current in this frame. To obtain this probability,
we consider the same measurement performed
in the laboratory frame $\mathcal L _0$, which becomes a time-shifted
measurement experiment and corresponds to the configuration
of figure~\ref{f2}. The probability of this
two time measurement in the laboratory frame
is then obtained by using the standard
Schr\"{o}dinger theory and the expression (\ref{4-a}) of
the state
vector, and of course state vector
reduction after the first measurement. Finally, we use the fact that probabilities are
relativistic
scalars to obtain the probability of the measurement as
seen
in the moving frame $\mathcal L _1$; from this infer
properties of the
probability current and bi-trajectories in this frame.

For the same reason as in \S~\ref{galilean},
if the detectors
are inserted in channel $A_2$ and $b_1$, the probability
vanishes, but does not if they are inserted in channel
 $A_2$ and $b_2$.
All bi-trajectories  that reach  $A_2$ therefore
have
 to pass through
$b_2$.  If we extend the (unperturbed) bi-trajectories
by continuity we see that, in Lorentz frame
$\mathcal L _1$, all bi-trajectories reaching outputs
$A_2$ and $B_2$ necessarily pass through $b_2$. Since,
as above, expression (\ref{2}) shows that no bi-trajectory
passes
through channels $a_2$ and $b_2$,  we conclude that
these bi-trajectories necessarily pass through $a_1$
and $b_2$.

Now consider a Lorentz frame $\mathcal L _2$ moving with
the opposite velocity $-v$ with respect to $\mathcal L _0$.
The same reasoning then shows that all
bi-trajectories reaching outputs
$A_2$ and $B_2$ have to pass through $a_2$
and $b_1$. The conclusion is that the bi-trajectories
reaching $A_2$ and $B_2$ have to look completely different
in frames $\mathcal L _1$ and$\mathcal L _2$: they
cannot be
obtained by merely applying a Lorentz transformation to each
 of their points. In short, the bi-trajectories cannot be relativistically
 invariant.

\section{Conclusion}

We therefore have a situation for a two particle system where, while the various
Lorentz observers
agree on the values of the probabilities of all
measurement
results, they may disagree on the path
taken by the probability in configuration space to reach
the detectors. Different motions of the probability fluid
in their respective frames are predicted by the theory.
Technically, while they agree when they calculate the flux of the probability
current (more precisely, its integral over time), they
disagree on the lines of current that lead to this flux.
The major difference
of points of view occurs between
observers
with positive velocity $v$ and those
with negative velocity. When $v$ vanishes,
a discontinuous change occurs, which is reminiscent of
the discontinuity
taking place in Galilean
relativity when the experimental setup is changed.
Nevertheless,
 a discontinuity of the description of the same
experiment by different observers raises a more
serious
problem, since it cannot be solved by invoking the Bohrian
wholeness of the experimental apparatus.
If one retains, as usual, the relativistic equivalence between all
Lorentz frames, it therefore seems difficult
to attribute physical reality to the probability current in configuration space, and
consequently to the probability fluid itself. In other words,
the theory tells us where and when the probability arrives,
but not how it propagates in space, which is not compatible with the motion of a fluid.

All our reasoning took place within standard quantum
mechanics,
but of course it has consequences if one uses the dBB
theory.
As pointed out in Ref.~\cite{Berndl-et-al},
the so called quantum equilibrium condition of this theory
cannot be relativistically invariant; see also
Refs. \cite{DGMZ, DGNSZ}. One possibility is
to
decide that the theory requires the
choice of a privileged frame, for instance that of the
laboratory
(but, even with this rule, problems arise when one
considers
experimental setups containing rapidly moving
components \cite{Gisin-2, Gisin-3}),
or the frame of the cosmic microwave radiation.

One may consider even less orthodox interpretations of quantum mechanics, and take another point of view emphasizing waves instead of point positions.
The dBB theory achieves two
main results: first, it unifies the dynamics, considering
that
measurements are just ordinary quantum interaction
processes,
and eliminating the need for any specific postulate
(von Neumann
projection, Heisenberg cut, etc.); second, it provides
a description   of
reality (an ontology) expressed in terms of point positions
 of particles. Clearly, the second result is incompatible
 with a completely relativistic theory\footnote{In the context
 of field theory, Bell
 \cite{Bell-beables} discusses why
 local occupation numbers (beables) for a fermionic field cannot
 provide a fully relativistic view of reality either.}, where no Lorentz
 frame is privileged. But, even if one gives up the description
 of reality with point particles, one can still preserve the first
 result: the positions are then not considered as real,
 but just as a mathematical indicator
 pointing to the branch of the state vector that is active,
 eliminating the empty waves. In this unorthodox
 \textquotedblleft reverse Bohmian interpretation\textquotedblright ,
 the unification of the dynamics
 is preserved, and the
 ultimate description of reality is obtained
 in terms of the active branch of the state vector. This view is reminiscent
 of Schr\"{o}dinger's initial hopes, considering the waves as real, while it requires to accept
 that  these waves propagate in configuration space (instead of ordinary space).
 It could also be seen as a variant of the Everett interpretation, but without many parallel
 versions of the world; this simplification eliminates the difficult definition of the
 probabilities of human perceptions in the many worlds, and may make the theory more acceptable.
 True, the
 indicator that determines this active branch is then
 not
 relativistically invariant; but, since it appears as a
 purely mathematical variable in the dynamics,
 this does not necessarily ruin
 the approach.

In conclusion, and whether or not a unified dynamics is
desirable
in quantum mechanics, it is clear that even blurred relativistic
trajectories seem to be incompatible with this theory -- this remains true
if they are blurred much more than what is requested by the
Heisenberg uncertainty relations. Bohr went even further
\cite{Bohr-Como} by claiming that,
due to the non-zero value of the Planck constant,
\textquotedblleft the space-time
description and the claims of causality are of complementary nature\textquotedblright ,
and therefore incompatible;
this should
indeed forbid a description of the motion of the
probability
fluid in space-time. Discontinuities in the trajectories taking
place when the Lorentz reference frame is changed
may be seen as a particular case of application of this
general statement.

%
%
%



\end{document}